\documentclass[journal]{IEEEtran}
\usepackage{graphicx}
\usepackage{subfigure}
\usepackage{amsmath}
\usepackage{amssymb}
\usepackage{latexsym}
\usepackage{url}
\usepackage{cite}
\usepackage{float}
\usepackage[T1]{fontenc}
\usepackage{authblk}
\usepackage{textcomp}
\usepackage{color}
\usepackage{siunitx}
\usepackage[utf8]{inputenc}
\usepackage{setspace}
\usepackage{amsmath,lipsum}
\usepackage{cuted} \stripsep-3pt
\newcommand*{\dif}{\mathop{}\!\mathrm{d}}
\makeatletter

\newcommand{\Rmnum}[1]{\expandafter\@slowromancap\romannumeral #1@}
\makeatother
\author{Xingjie Fan, Dawei Wang,~\IEEEmembership{Member,~IEEE,} Julian Cheng,~\IEEEmembership{Senior Member,~IEEE,} Jingkai Yang, \\and Jing Ma,~\IEEEmembership{Member,~IEEE}
\thanks{Xingjie Fan, Jingkai Yang and Jing Ma are with National Key Laboratory of Tunable Laser Technology, Harbin Institute of Technology, Harbin, China (e-mail: fanxingjie@hit.edu.cn; ynyxyjk@gmail.com; majing@hit.edu.cn).}
\thanks{Dawei Wang is with School of Electronics and Information, Northwestern Polytechnical University, Xi’an, China (e-mail: wangdw@nwpu.edu.cn).}
\thanks{Julian Cheng is with School of Engineering, The University of British Columbia, Kelowna, BC, Canada (e-mail: julian.cheng@ubc.ca).}

}

\begin{document}
\title{Few-Mode Fiber Coupling Efficiency for Free-Space Optical Communication}
\maketitle

\begin{abstract}
Few-mode fiber is a significant component of free-space optical communication at the receiver to obtain achievable high coupling efficiency. A theoretical coupling model from the free-space optical communication link to a few-mode fiber is proposed based on a scale-adapted set of Laguerre-Gaussian modes. It is found that the coupling efficiency of various modes behaves differently in the presence of atmospheric turbulence or random jitter. Based on this model, the optimal coupling geometry parameter is obtained to maximize the coupling efficiency of the selected mode of few-mode fiber. The communication performance with random jitter is investigated. It is shown that the few-mode fiber has better bit-error rate performance than single-mode fiber, especially in high signal-to-noise ratio regimes.
\end{abstract}
\section{INTRODUCTION}
\IEEEPARstart{F}{ree-space} optical communication (FSO) has attracted extensive attention for the advantages of large bandwidth, low cost, and flexible implementation \cite{boroson2014overview,leitgeb2005optical}. Compared with conventional radio frequency systems, FSO is more susceptible to external factors, including atmospheric turbulence and random jitter\footnote{Random jitter is the jitter that is due to random vibrations of the optical platform mount, which cause the variation of the optical signal at the receiver.}\cite{10.1117/12.864274,yarnall2017analysis}. Both turbulence and jitter tend to reduce the coupling efficiency from the free-space beam to fiber and lead to severe power fading and scintillation. The loss of receiving power deteriorates the instantaneous signal-to-noise ratio (SNR), resulting in an increase in the average bit-error rate (BER) \cite{dikmelik2005fiber}. For mitigating the impact of external factors on communication performance, a series of single-mode fiber (SMF) based technologies have been adopted, such as high power erbium-doped fiber amplifier (EDFA) and coherent receiver\cite{Song:s,Zhang:s}.

With the advancement of mode multiplexing technology such as photonic lantern multiplexer\cite{leon2013photonic}, coupling from the FSO link to few-mode fiber (FMF) can be compatible with current SMF based optical communication systems\cite{zheng2018performance}. Besides, FMF has a larger core diameter, which makes it easier to adjust the focus process of the FSO system. In addition, unlike SMF that only supports fundamental mode, FMF can also support several high-order modes caused by turbulence and jitter. FMF's excellent performance can increase achievable fiber coupling efficiency in the presence of turbulence or jitter. Specifically, compared with the SMF, the coupling efficiencies for a three-mode FMF and a six-mode FMF are improved by $\sim$4dB and $\sim$7dB in the presence of turbulence\cite{zheng2016free}. In a follow-up work, the power fluctuation of the FSO receiver can be reduced by 5$\sim$11dB under moderate to strong turbulence with mode diversity coherent reception implemented by FMF coupling, mode de-multiplexing, and offline digital signal processing\cite{zheng2018performance}. Besides, FMF can be combined with a fast-steering mirror control loop\cite{geisler2018experimental} or a multi-aperture multi-spatial-mode receiver \cite{geisler2018ground} to mitigate atmospherically induced tilt.

However, most of the above studies draw conclusions on coupling efficiency based on experimental results and focus on the applications of FMF in FSO. There is a lack of theoretical foundation for evaluating the coupling efficiency from the FSO link to FMF. A simple and effective coupling model can be used to analyze the variation trend of the coupling efficiency of each mode under different conditions such as turbulence and random jitter. We can adjust the coupling efficiency of the selected modes by optimizing relevant parameters to achieve a specific power distribution arrangement. Due to formidable mathematical complexity, it is challenging to obtain an accurate solution of the coupling efficiency directly from the linearly polarized (LP) mode of a step-index fiber. To the best of the authors' knowledge, only Alireza Fardoost \emph{et al.} provided an approximate theoretical coupling model \cite{fardoost2019optimizing}. However, this model only considers the free-space modes as the forms of Hermite–Gaussian (HG) or Laguerre–Gaussian (LG) beams, which is idealistic and cannot illustrate the effects of atmospheric turbulence or random jitter on coupling efficiency.

We propose a theoretical coupling model from free-space to FMF in the presence of atmospheric turbulence or random jitter\footnote{Aberration is also an important factor affecting the coupling efficiency, especially in astronomy and satellite-based FSO applications. For more common terrestrial application scenarios, we ignore this factor in our research.}, where the LP mode of a step-index fiber is approximated by a scale-adapted set of LG modes \cite{bruning2015overlap}. As a comparison, the coupling efficiency of the $\mathrm{LP_{01}}$ mode can be approximated as the coupling efficiency of SMF. We approximate the incident optical field as a plane wave in the presence of atmospheric turbulence. Alternatively, we approximate the incident optical field as a Gaussian beam to consider random jitter. Since atmospheric turbulence and random jitter have different effects on the coupling efficiency, and these two external effects have different mathematical approximations, we discuss these two cases separately. In our research, the incident light field is assumed to be excited by a single-mode laser. Therefore, after propagating through free-space, whether the incident light field is approximated as a plane wave or a Gaussian beam, the fundamental mode is regarded as the essential component of the incident light field. Moreover, based on the variation trend of the coupling efficiency of each mode, we maximize the coupling efficiency of FMF by adjusting the relevant parameters. Finally, we discuss communication performance in the presence of random jitter.

This paper is organized as follows. Section \Rmnum{2} presents theoretical formulations. Section \Rmnum{3} provides numerical results and analysis. Section \Rmnum{4} concludes the paper.

\section{THEORETICAL FORMULATIONS}
\begin{figure}[tbp]
\centering
\includegraphics[width=2.5in]{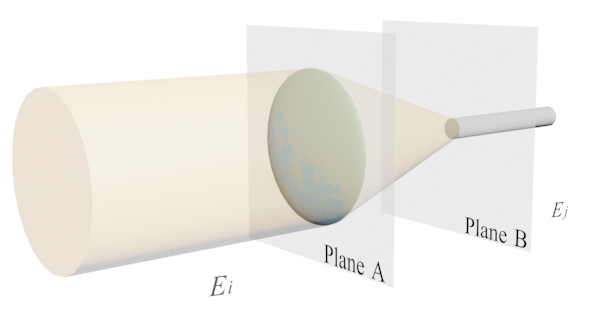}
\caption{A thin lens couples incident optical field $\mathbf{E_i}$ into FMF (the mode field is $E_{j,B}$), whose end face lies on plane $B$; the aperture plane is denoted by plane $A$.}
\end{figure}

The fiber coupling efficiency is defined as the ratio of the average power coupled into the fiber, $\left<\mathbf{P_{c}}\right>$, to the average available power in the receiver aperture plane, $\left<\mathbf{P_a}\right>$, where the angle brackets represent the ensemble-average operator; bold type indicates random quantities. The coupling efficiency of the $j$-th mode is defined as \cite{winzer1998fiber,dikmelik2005fiber}
\begin{equation}
\begin{aligned}
\eta_j
=\frac{\left<\mathbf{P_{c,j}}\right>}{\left<\mathbf{P_a}\right>}
=\frac{\left<|\int_B\!\mathbf{E_{i,B}}(\vec r)E_{j,B}^*(\vec r)\dif\vec{r}|^2\right>}{\left<\int_A\!|\mathbf{E_{i,A}}(\vec r)|^2\dif\vec r\right>}.
\label{eta_j}
\end{aligned}
\end{equation}
In Fig. 1, $\mathbf{E_{i,A}}(\vec r)$ and $\mathbf{E_{i,B}}(\vec r)$ characterize the incident random optical field upon plane $A$ and plane $B$; $E_{j,B}^*(\vec r)$ is the complex conjugate of the $j$-th normalized fiber mode field. Both integrals in the numerator and denominator of (\ref{eta_j}) can evaluate on the plane $A$ because it is more convenient. Assuming the average input optical intensity is independent of $\vec r$, eq. (\ref{eta_j}) can be rewritten as
\begin{equation}
\begin{aligned}
\eta_j=
\frac{4}{\pi d_R^2} \int\!\!\!\int_A\!\mu_i(\vec{r}_1,\vec{r}_2)E_{j,A}^*(\vec{r}_1)E_{j,A}(\vec{r}_2)\dif\vec{r}_1\dif\vec{r}_2
\label{etaj}
\end{aligned}
\end{equation}
where $d_R$ is the receiver lens diameter; $E_{j,A}(\vec{r})$ is the backpropagated fiber mode field on plane A; the mutual coherence function of the incident optical field is given by
\begin{equation}
\begin{aligned}
\mu_i(\Vec{r}_1,\Vec{r}_2)=
\frac{\left<\mathbf{E_{i,A}}(\Vec{r}_1)\mathbf{E_{i,A}^*}(\Vec{r}_2)\right>}{\left[\left<|\mathbf{E_{i,A}}(\Vec{r}_1)|^2\right>\left<|\mathbf{E_{i,A}}(\Vec{r}_2)|^2\right>\right]^{1/2}}.
\label{mu_i}
\end{aligned}
\end{equation}

For a weakly guided step-index FMF, the field distribution of the guided modes $E_{j,B}(\vec r)$ can be represented by the solution to the scalar Helmholtz equation, also known as LP modes. Since the step-index fiber has cylindrical symmetry, and this cylindrical symmetry can be also defined as LG modes, which are the solutions to the paraxial Helmholtz equation in a cylindrically symmetric coordinate system. Therefore, we use the LG modes to approximate the LP modes. The solution of LG modes at the waist position can be represented as 
\begin{equation}
\begin{aligned}
\mathrm{LG}_{pl}(r,\phi)=\frac{B_{pl}}{\omega_0}\left(\sqrt{2}\frac{r}{\omega_0}\right)^{l}L^{l}_p\left(\frac{2r^2}{\omega^2_0}\right)\exp\left({-\frac{r^2}{\omega_0^2}}\right)e^{-il\phi}
\label{LG}
\end{aligned}
\end{equation}
where $l$ and $p$ are the indices for the guided azimuthal and radial components; $L^{l}_p$ are the associated Laguerre polynomials; $\omega_0$ is the fundamental Gaussian radius; $B_{pl}=\left(\frac{2p!}{\pi(l+p)!}\right)^\frac{1}{2}$ is a normalization factor.
For both LP modes and LG modes, they have similar azimuthal and radial dependence with the same independent variables. Therefore, corresponding modes can be found by choosing the same azimuthal order, and the field functions with the same amount of roots in the radial direction. A scale parameter $h_{lp}$ is defined to evaluate the best possible matching of both mode sets. The matching of the modes can be presented by the overlap relation
\begin{equation}
\begin{aligned}
h_{lp}=
\int\!\!\!\int_B\! \mathrm{LP}_{lp}(r,\phi)\mathrm{LG}^*_{p-1,l}(r,\phi)\dif r\dif\phi
\end{aligned}
\end{equation}
where $\mathrm{LP}_{lp}(r,\phi)$ is the LP mode. After normalization for both mode sets, the value can vary between $h_{lp}=0$ for orthogonal fields, and $h_{lp}=1$ for perfectly matched fields. As the overlap value $h_{lp}$ increases, a better approximation can be achieved. It is found that the LP modes have to be far from their cutoff condition and of low radial order, to provide high overlap relation \cite{bruning2015overlap}. Specifically, as a six-mode FMF mentioned in their research, the overlap relation can reach $h_{lp}=0.99$ for $\mathrm{LP_{01}}$, $\mathrm{LP_{11}}$ and $\mathrm{LP_{21}}$ modes, and $h_{lp}=0.98$ for $\mathrm{LP_{02}}$ mode, by adjusting the beam-to-core radius ratio. The beam-to-core radius ratio is defined as a scale factor from LG mode beam radius to fiber core radius, and we can use this ratio to define $\omega_0$ when the fiber core radius is determined. Besides, when the V parameter of FMF increases and moves away from the cutoff of a low radial order mode, the overlap relation of that mode satisfies $0.94<h_{lp}<1$ for a wide range of V parameter values\cite{bruning2015overlap}, and thus we no longer discuss the effect of the V parameter on coupling efficiency. Therefore, we use (\ref{LG}) to represent the electric field distribution of the guided modes $E_{j,B}(\vec r)$ in FMF.
In our research, we ignore the correlation terms between modes, and the total coupling efficiency can be expressed as
\begin{equation}
\begin{aligned}
\eta_{tot}=\sum_{j=1}^n\eta_j
\label{etatot}
\end{aligned}
\end{equation}
where $n$ is the total number of modes held by the FMF. Since atmospheric turbulence and random jitter have different effects on the coupling efficiency, and these two external effects have different mathematical approximations, we next discuss these two cases separately.
\subsection{Fiber Coupling Efficiency in the Presence of Atmospheric Turbulence}

The mutual coherence function of a plane wave distorted by atmospheric turbulence can be expressed in the form of Gaussian function as \cite{dikmelik2005fiber}
\begin{equation}
\begin{aligned}
\mu_{i}^{\scriptscriptstyle T}(\vec{r}_1,\vec{r}_2)
=\exp\left(-|\vec{r}_1-\vec{r}_2|^2/\rho^2\right)
\label{muit}
\end{aligned}
\end{equation}
where the superscript $T$ represents the presence of atmospheric turbulence; $\rho=\left(1.46C_n^2k^2L\right)^{-3/5}$ is the spatial coherence distance that assumes a Kolmogorov power-law spectrum for the refractive-index fluctuations\cite{andrews2005laser}; $C_n^2$ is the refractive-index structure constant; $k$ is the wavenumber of the optical field; and $L$ is the communication link distance.

We assume $E_{j,A}(\vec r)$ has the similar expression as $E_{j,B}(\vec r)$, and we use $\omega$ to denote the radius of backpropagated fiber mode at plane $A$, which plays a similar role to $\omega_0$ on plane B. This assumption is widely used in SMF coupling efficiency calculations\cite{winzer1998fiber,dikmelik2005fiber}. Substituting (\ref{LG}) and (\ref{muit}) into (\ref{etaj}), we write the coupling efficiency of the $j$-th mode under atmospheric turbulence as
\begin{equation}
\begin{aligned}
\eta_j^{\scriptscriptstyle T}&=\frac{4B^2_{pl}}{\pi\omega^2d_R^2}\int_0^{d_R/2}\!\!\!\int_0^{d_R/2}\!\!\!\int_0^{2\pi}\!\!\int_0^{2\pi}\!L^{l}_p\left(\frac{2r_1^2}{\omega^2}\right)L^{l}_p\left(\frac{2r_2^2}{\omega^2}\right)\\
&\times\,\exp\left[-\frac{\left(r_1^2+r_2^2-2r_1r_2\cos(\phi_1-\phi_2)\right)}{\rho^2}\right]
\left(\frac{2}{\omega^2}r_1r_2\right)^l\\
&\times\,\exp\left(-\frac{r_1^2+r_2^2}{\omega^2}\right)e^{-il(\phi_1-\phi_2)}r_1r_2\dif\phi_1\dif\phi_2\dif r_1\dif r_2.
\label{etajT}
\end{aligned}
\end{equation}
In (\ref{etajT}), the double integral over the angle variables $\phi_1$ and $\phi_2$ that needs to be evaluated first and is given by
\begin{equation}
\begin{aligned}
I=\int_0^{2\pi}\!\!\!\int_0^{2\pi}\!\exp\left[\frac{2r_1r_2\cos(\phi_1-\phi_2)}{\rho^2}\right]e^{-il(\phi_1-\phi_2)}\dif\phi_1\dif\phi_2.
\end{aligned}
\end{equation}
Making a change of variables to $\phi=\phi_1-\phi_2$ and $\phi_a=\phi_2$ to evaluate the integral over $\phi_a$ yields. After some algebraic manipulations, we arrive
\begin{equation}
\begin{aligned}
I=4\pi^2I_{l}\left(\frac{2r_1r_2}{\rho^2}\right)
\label{IT}
\end{aligned}
\end{equation}
where $I_l$ denotes the $l$th-order modified Bessel function of the first kind. We next normalize the radial integration variables to the receiver lens radius and define $x_1=2r_1/d_R$ and $x_2=2r_2/d_R$. Substituting the result for the integral $I$ given by (\ref{IT}) into the coupling efficiency expression of (\ref{etajT}), we obtain
\begin{equation}
\begin{aligned}
\eta_j^{\scriptscriptstyle T}&=2\pi B^2_{pl}\int_0^{1}\!\!\!\int_0^{1}\!I_{l}\left(\frac{A_R}{A_C}2x_1x_2\right)\exp\left(-\frac{A_R}{A_C}(x_1^2+x_2^2)\right)\\
&\times\,L^{l}_p\left(2\gamma^2x_1^2\right)L^{l}_p\left(2\gamma^2x_2^2\right)\\
&\times\,\left(2\gamma^2x_1x_2\right)^{l+1}\exp\left(-\gamma^2(x_1^2+x_2^2)\right)\dif x_1\dif x_2.
\label{etajT2}
\end{aligned}
\end{equation}
Here $A_R=\pi d_R^2/4$ is the aperture area\footnote{The incident light field is assumed to completely cover the aperture.}; $A_C=\pi\rho^2$ is the spatial coherence area of the incident plane wave $\mathbf{E_{i,A}}(\vec r)$, also known as speckle size. The ratio $A_R/A_C$ represents the number of speckles over the receiver aperture area, which is equivalent to the ratio of beam diameter to atmospheric coherence length $d_R/\rho$, and we can use this ratio to represent the turbulence strength\cite{zheng2018performance}. The larger the ratio $A_R/A_C$ is, the stronger the turbulence becomes, and vice versa. Eq. (\ref{etajT2}) also shows the coupling efficiency of the $j$-th mode in the FMF depends on the coupling geometry through a single parameter $\gamma$ given by
\begin{equation}
\begin{aligned}
\gamma=\frac{d_R}{2\omega}.
\label{gamma}
\end{aligned}
\end{equation}
This parameter is the ratio of the receiver lens radius to the radius of the backpropagated fiber mode at the lens. In the absence of atmospheric turbulence, when {$A_R/A_C\!\rightarrow\!0$}, eq. (\ref{etajT2}) can be rewritten as
\begin{equation}
\begin{aligned}
\eta_j=
\left\{
       \begin{array}{lr}
       8\gamma^2\left|\int_0^{1}\!L_p\left(2\gamma^2x^2\right)x\exp\left(-\gamma^2x^2\right)\dif x\right|^2, &l=0,\\
       0\,,  &l\neq0.
       \end{array}
\right.
\label{t0}
\end{aligned}
\end{equation}
When $\gamma=1.12$, the maximum coupling efficiency of the $\mathrm{LP_{01}}$ mode calculated from (\ref{t0}) is 0.81, it agrees with the result obtained in previous studies\cite{winzer1998fiber,dikmelik2005fiber}. As we ignore the correlation terms between modes, the total coupling efficiency under the influence of atmospheric turbulence can be obtained by substituting (\ref{etajT2}) into (\ref{etatot}).

\subsection{Fiber Coupling Efficiency in the Presence of Random Jitter}
To facilitate the analysis of the effect of random jitter on coupling efficiency, we assume the initial incident beam has a Gaussian intensity distribution\cite{Imai:75}. The normalized amplitude distribution of the electric field at distance $z$ is expressed as
\begin{equation}
\begin{aligned}
M(\Vec{r},z)=\sqrt{\frac{2}{\pi\omega_z^2}}\exp\left(-\frac{r^2}{\omega_z^2}-ik\frac{r^2}{2R_z}-ikz+i\zeta(z)\right)
\label{M}
\end{aligned}
\end{equation}
where $\omega_z=\omega_G\sqrt{1+z/Z_0}$ is the radius at which the field amplitudes fall to $1/e$ of their axial values at $z$ along the beam; $Z_0=\pi\omega^2_G/\lambda$ is Rayleigh range; $\omega_G$ is the waist radius of the beam; $\lambda$ is the wavelength; $R_z=z\left[1+\left(Z_0/z\right)^2\right]$ is the radius of curvature of the beam's wavefronts at $z$; and $\zeta(z)$ is the Gouy phase at $z$, an extra phase term beyond that attributable to the phase velocity of light. We define $d$ as the static radial offset of the optical beam from the nominal axis of the lens (plane $A$). Without loss of generality, if there is an offset bias $d$, we can assume $\Vec{d}$ is located along the same direction with $\Vec{r}$. Then the distribution concerning the $\Vec{r}$ and $\Vec{d}$ is given by 
\begin{equation}
\begin{aligned}
M(\Vec{r},\Vec{d},z)&=M(r,\theta,d,z)\\
&=\sqrt{\frac{2}{\pi\omega_z^2}}\exp\left(-ikz+i\zeta(z)\right)\\
&\times\,\exp\left[-\left(\frac{1}{\omega_z^2}+\frac{ik}{2R_z}\right)(r^2-2rd\mathrm{cos}\theta+d^2)\right].
\label{Md}
\end{aligned}
\end{equation}

Therefore, the mutual coherence function of a Gaussian beam with an offset bias $d$ can be expressed by substituting (\ref{M}) and (\ref{Md}) into (\ref{mu_i})
\begin{equation}
\begin{aligned}
\mu_{i}^{\scriptscriptstyle J}(\vec{r}_1,\vec{r}_2)&=
\frac{M(\vec{r}_1,\Vec{d},z)M^*(\vec{r}_2,\Vec{d},z)}{\left(|M(\vec{r}_1,z)|^2|M(\vec{r}_2,z)|^2\right)^{1/2}}\\
&=\exp\left[-\frac{ik}{2R_z}(r_1^2-r_2^2)+\frac{2d}{\omega_z^2}(r_1\mathrm{cos}\theta_1+r_2\mathrm{cos}\theta_2)\right]\\
&\times\,\exp\left[-\frac{2d^2}{\omega_z^2}+\frac{ikd}{R_z}(r_1\mathrm{cos}\theta_1-r_2\mathrm{cos}\theta_2)\right]
\label{muij}
\end{aligned}
\end{equation}
where the superscript $J$ represents the presence of random jitter. Substituting (\ref{LG}) and (\ref{muij}) into (\ref{etaj}), we express the coupling efficiency of the $j$-th mode with an offset bias $d$ between the center of the focused beam and the nominal axis of the lens as
\begin{equation}
\begin{aligned}
\eta_j^{\scriptscriptstyle J}(d)
&=\frac{4B^2_{pl}}{\pi\omega^2d_R^2}\int_0^{d_R/2}\!\!\!\int_0^{d_R/2}\!\!\!\int_0^{2\pi}\!\!\int_0^{2\pi}\!\exp\left[-\frac{ik}{2R_z}(r_1^2-r_2^2)\right]\\&\times\,
\exp\left[\frac{ikd}{R_z}(r_1\cos\theta_1-r_2\mathrm{cos}\theta_2)-il(\theta_1-\theta_2)\right]\\&\times\,
\exp\left[\frac{2d}{\omega_z^2}(r_1\mathrm{cos}\theta_1+r_2\mathrm{cos}\theta_2)-\frac{2d^2}{\omega_z^2}-\frac{r_1^2+r_2^2}{\omega^2}\right]\\&\times\,
\left(\frac{2}{\omega^2}r_1r_2\right)^lL^{l}_p\left(\frac{2r_1^2}{\omega^2}\right)L^{l}_p\left(\frac{2r_2^2}{\omega^2}\right)\\&\times\,
r_1r_2\,\dif\theta_1\dif\theta_2\dif r_1\dif r_2.
\label{etajJ}
\end{aligned}
\end{equation}
Similarly, the double integral over the angle variables $\theta_1$ and $\theta_2$ needs to be evaluated first. Then, making a change of variables to $\theta=\theta_1-\theta_2$ and $\theta_a=\theta_2$, we evaluate the integral over $\theta_a$ to obtain the integral result. After some algebraic manipulations, we arrive at
\begin{equation}
\begin{aligned}
K&=\int_0^{2\pi}\!\!\int_0^{2\pi}\!\exp\left[\frac{2d}{\omega_z^2}(r_1\mathrm{cos}\theta_1+r_2\mathrm{cos}\theta_2)\right]\\
&\times\,\exp\left[\frac{ikd}{R_z}(r_1\mathrm{cos}\theta_1-r_2\mathrm{cos}\theta_2)-il(\theta_1-\theta_2)\right]\dif\theta_1\dif\theta_2 \\
&=4\pi^2I_l\left(\frac{2dr_1}{\omega_z^2}+\frac{kdr_1}{R_z}i\right)I_l\left(\frac{2dr_2}{\omega_z^2}-\frac{kdr_2}{R_z}i\right).
\label{K}
\end{aligned}
\end{equation}
We next normalize the radial integration variables to the receiver lens radius and define $x_1=2r_1/d_R$ and $x_2=2r_2/d_R$. Substituting the result for the integral $K$ given by (\ref{K}) into the coupling efficiency expression of (\ref{etajJ}), we obtain
\begin{equation}
\begin{aligned}
\eta_j^{\scriptscriptstyle J}(d)&=2\pi B^2_{pl}\exp\left(-\frac{2d^2}{\omega_z^2}\right)\left(2\gamma^2\right)^{l+1}\\
&\times\,\left|\int_0^1\!I_l\left(\alpha xd\right)\exp\left(-\beta x^2\right)x^{l+1}L^{l}_p\left(2\gamma^2x^2\right)\dif x\right|^2
\label{etajJ2}
\end{aligned}
\end{equation}
where $\alpha=\frac{d_R}{\omega_z^2}+\frac{kd_R}{2R_z}i;\beta=\gamma^2+\frac{kd_R^2}{8R_z}i.$ Similarly, in the absence of offset bias, i.e., $d=0$, eq. (\ref{etajJ2}) can be rewritten as 

\begin{equation}
\begin{aligned}
\eta_j=
\left\{
       \begin{array}{lr}
       8\gamma^2\left|\int_0^{1}\!L_p\left(2\gamma^2x^2\right)x\exp\left(-\beta x^2\right)\dif x\right|^2,\quad &l=0,\\
    0\,,&l\neq0.
       \end{array}
\right.
\label{d0}
\end{aligned}
\end{equation}
If $R_z\!\rightarrow\!\infty$, in which case the incident beam can be regarded as a plane wave, the equation $\beta=\gamma^2$ is satisfied, and (\ref{d0}) would have the same expression as (\ref{t0}).

Consider independent identical Gaussian distributions for the elevation and the horizontal displacement\cite{arnon2003effects}. The radial displacement $d$ at the receiver is modeled by a Rayleigh distribution
\begin{equation}
\begin{aligned}
p(d)=\frac{d}{\sigma^2}\exp\left(-\frac{d^2}{2\sigma^2}\right).
\end{aligned}
\end{equation}
where $\sigma^2$ is the jitter variance at the receiver. Therefore, integrating the random variable $d$, we obtain the expected value of the coupling efficiency of the $j$-th mode for the random jitter as
\begin{equation}
\begin{aligned}
\left<\eta_j^{\scriptscriptstyle J}\right>&=2\pi B^2_{pl}\left(2\gamma^2\right)^{l+1}\!\!\int_0^{\infty}\!\exp\left(-\frac{2d^2}{\omega_z^2}\right)\frac{d^2}{\sigma^4}\exp\left(-\frac{d^2}{\sigma^2}\right)\\&\times\,
\left|\int^1_0\!I_l\left(\alpha xd\right)\exp\left(-\beta x^2\right)x^{l+1}L^{l}_p\left(2\gamma^2x^2\right)\dif x\right|^2\dif d.
\label{etajJsigma}
\end{aligned}
\end{equation}
Using the properties of Bessel functions and an integral identity \cite[(6.631.7)]{gradshteyn2014table}, we can simplify (\ref{etajJsigma}) as
\begin{equation}
\begin{aligned}
\left<\eta_j^{\scriptscriptstyle J}\right>
&=\frac{\pi^2B^2_{pl}}{32\sigma^4\kappa^3(\sigma)}
\left(2\gamma^2\right)^{l+1}\\&\times\, 
\left|\int_0^1\!\exp\left[\frac{\alpha^2x^2}{8\kappa(\sigma)}-\beta x^2\right]\alpha x^{l+2}L^{l}_p\left(2\gamma^2x^2\right)I(x)\dif x\right|^2
\label{etajJsigma2}
\end{aligned}
\end{equation}
where $\kappa(\sigma)=\frac{1}{\omega_z^2}+\frac{1}{2\sigma^2}$; $I(x)=I_{\frac{1}{2}l-\frac{1}{2}}\left(\frac{-\alpha^2x^2}{8\kappa(\sigma)}\right)-I_{\frac{1}{2}l+\frac{1}{2}}\left(\frac{-\alpha^2x^2}{8\kappa(\sigma)}\right)$. If the link distance $z$ is large enough, i.e., $R_z\!\rightarrow\!\infty$, in which case the incident Gaussian beam can be regarded as a plane wave. We set $d_R=8\,\mathrm{cm}$ and $\omega_z/\omega=1$. In the absence of random jitter, when $\sigma\!\rightarrow\!0$, the maximum coupling efficiency of the $\mathrm{LP_{01}}$ mode is 0.81 with $\gamma=1.12$. It is also verified in previous studies\cite{winzer1998fiber,dikmelik2005fiber}.

Similarly, as we ignore the correlation terms between modes, the total coupling efficiency in the presence of random jitter can be obtained by substituting (\ref{etajJsigma2}) into (\ref{etatot}).

\subsection{Average Bit Error Rate due to Random Jitter}

The ratio $ A_R/A_C $ that represents the turbulence strength is a constant in a specific turbulent case. However, the offset bias $d$ is a random quantity obeying the Rayleigh distribution in a specific random jitter case. Therefore, we only analyze the effect of coupling efficiency on communication performance in the presence of random jitter.

The received optical power stimulates the photonic current of the photodetector; the SNR is represented by the parameter $Q$. The BER of an optical receiver for an intensity-modulation and direct-detection system with non-return-to-zero is given by\cite{toyoshima2006maximum}
\begin{equation}
\begin{aligned}
\mathrm{BER}(Q)=\frac{1}{2}\mathrm{erfc}\left(\frac{Q}{\sqrt{2}}\right)
\end{aligned}
\end{equation}
where erfc is the complementary error function. Therefore, the unconditional BER in the presence of jitter is averaged with respect to the probability density function of the received optical intensity and is given by
\begin{equation}
\begin{aligned}
\mathrm{BER}&=\int p(d)\mathrm{BER}\left[Q\ \eta_j^{\scriptscriptstyle J}(d)\right]\mathrm{d}d\\
&=\frac{1}{2}\int p(d)\mathrm{erfc}\left[\frac{Q}{\sqrt{2}}\eta_j^{\scriptscriptstyle J}(d)\right]\mathrm{d}d.\\
\end{aligned}
\end{equation}

\section{NUMERICAL RESULTS AND DISCUSSIONS}

\begin{figure}[tbp]
\centering
\subfigure[]{
\centering
\includegraphics[width=3.5in]{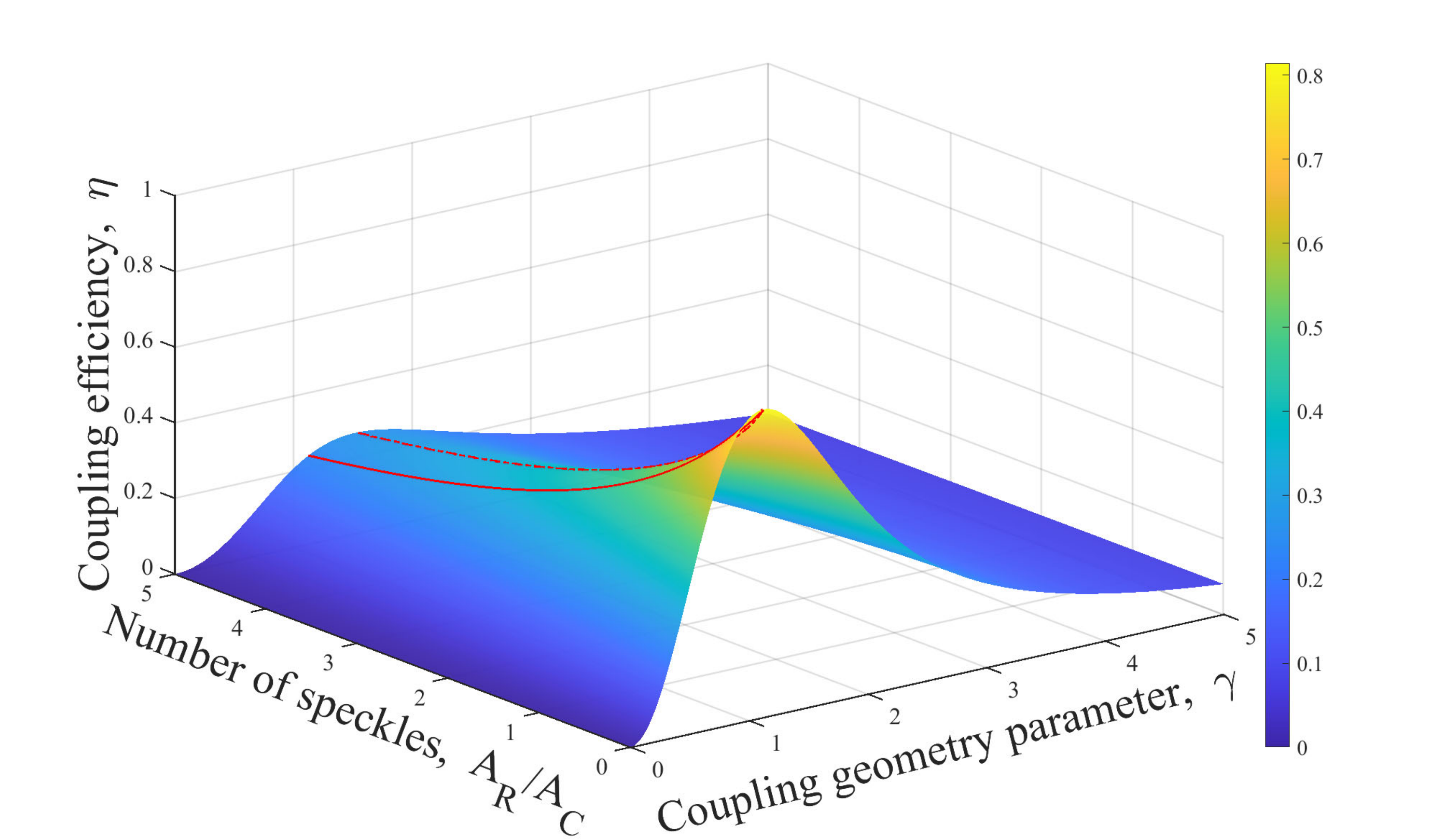}
}
\subfigure[]{
\centering
\includegraphics[width=3.5in]{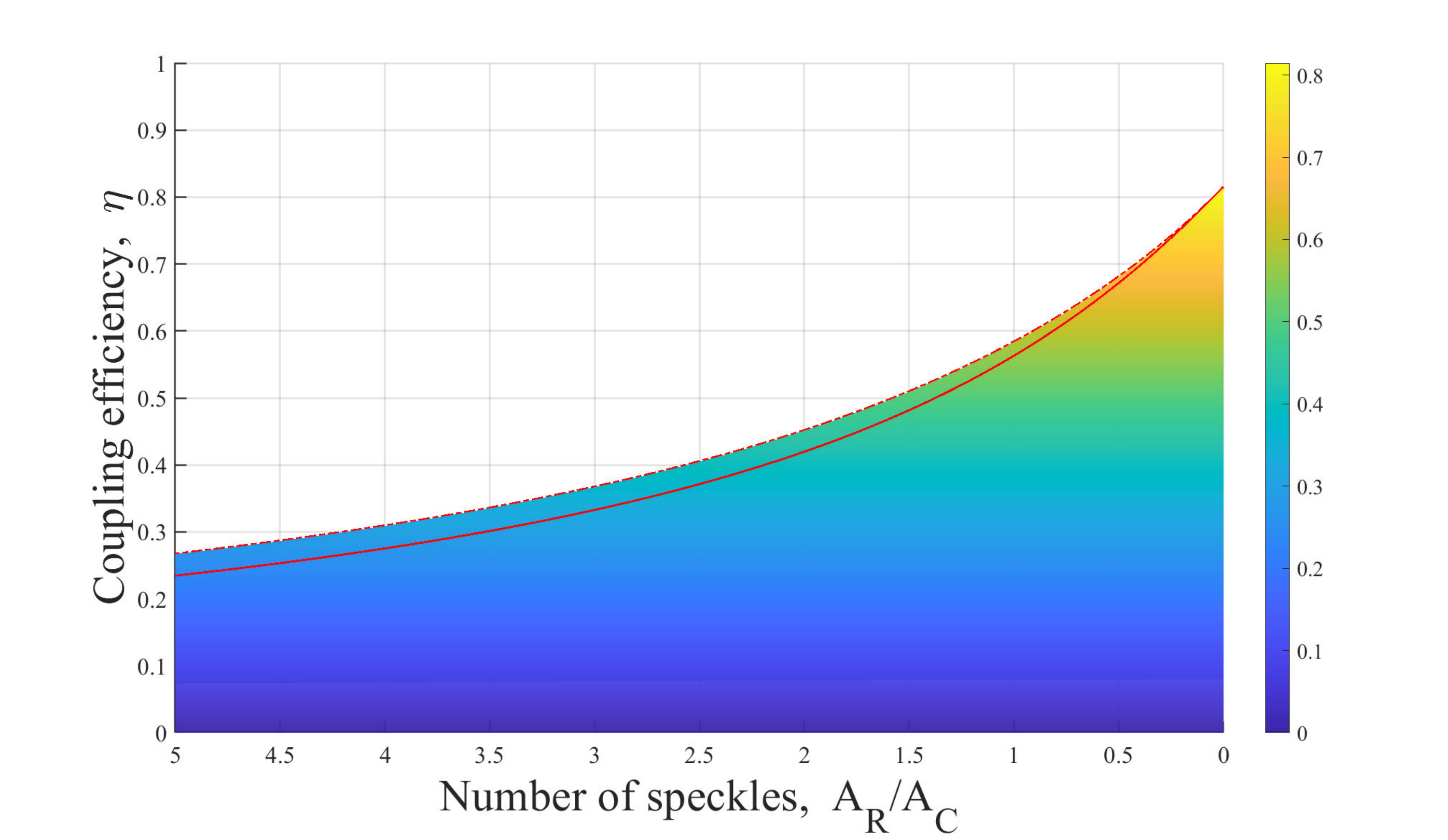}
}

\caption{Simulation results of the combined effects of the number of speckles $A_R/A_C$ and the coupling geometry parameter $\gamma$ on total coupling efficiency from free-space to a three-mode FMF. The red dashed curve is calculated for the varying value of $\gamma_3$, with the uniform change of $A_R/A_C$, the maximum value that $\eta_{tot}$ can achieve. The red solid curve represents when $\gamma_3\!=\!1.12$, $\eta_{tot}$ as a function of $A_R/A_C$. (a) is the main view and (b) is the left view.}
\end{figure}

In this section, we use three-mode FMF (holding $\mathrm{LP_{01}}$ and $\mathrm{LP_{11}}$ modes) and six-mode FMF (holding $\mathrm{LP_{01}}$, $\mathrm{LP_{11}}$, $\mathrm{LP_{21}}$ and $\mathrm{LP_{02}}$ modes) as representatives. We first obtain the optimum value of parameter $\gamma$, which is defined in (\ref{gamma}), to maximize the coupling efficiency of FMF in the presence of atmospheric turbulence. Then we present the numerical results of free-space to FMF coupling efficiency in the presence of atmospheric turbulence and random jitter. As a comparison, the coupling efficiency of the $\mathrm{LP_{01}}$ mode can be approximated as the coupling efficiency of SMF. Finally, we discuss communication performance under random jitter conditions.
\subsection{Optimum Value of Parameter $\gamma$ in the Presence of Atmospheric Turbulence}
It can be known from (\ref{etajT2}), the optimum value of the parameter $\gamma$ that maximizes the coupling efficiency depends on the number of speckles $A_R/A_C$ in the presence of atmospheric turbulence. Fig. 2 shows the combined effects of $A_R/A_C$ and $\gamma$\footnote{To simplify the analysis, we assume the radius of different backpropagated fiber modes are the same.} on the total coupling efficiency of a three-mode FMF, and we use $\gamma_3$ to represent the coupling geometry parameter of the three-mode FMF. In the limit of a deterministic optical plane wave incident upon the coupling lens (in the absence of turbulence, where $A_C\!\rightarrow\!\infty$), $\gamma_3\!=\!1.12$ is obtained where only the $\mathrm{LP}_{01}$ mode is excited \cite{snyder1969excitation}, i.e., we can consider it as an SMF in this condition. This value agrees with the result obtained in a previous study\cite{winzer1998fiber}. 

However, the optimum value of $\gamma_3$ varies with the change of $A_R/A_C$. As visualized in Fig. 2, when the ratio $A_R/A_C=5$ represents a strong turbulent condition, the difference in coupling efficiency for $\gamma_3=1.12$ and other optimum $\gamma_3$ reaches the maximum value of 0.033. We can find that when $\gamma_3\!=\!1.12$, the total coupling efficiency of a three-mode FMF is not appreciably less than the maximum coupling efficiency that can be obtained by other optimum $\gamma_3$. Therefore, we use $\gamma_3=1.12$ to calculate the coupling efficiency of a three-mode FMF.

\begin{figure}[tbp]
\centering
\includegraphics[width=2.5in]{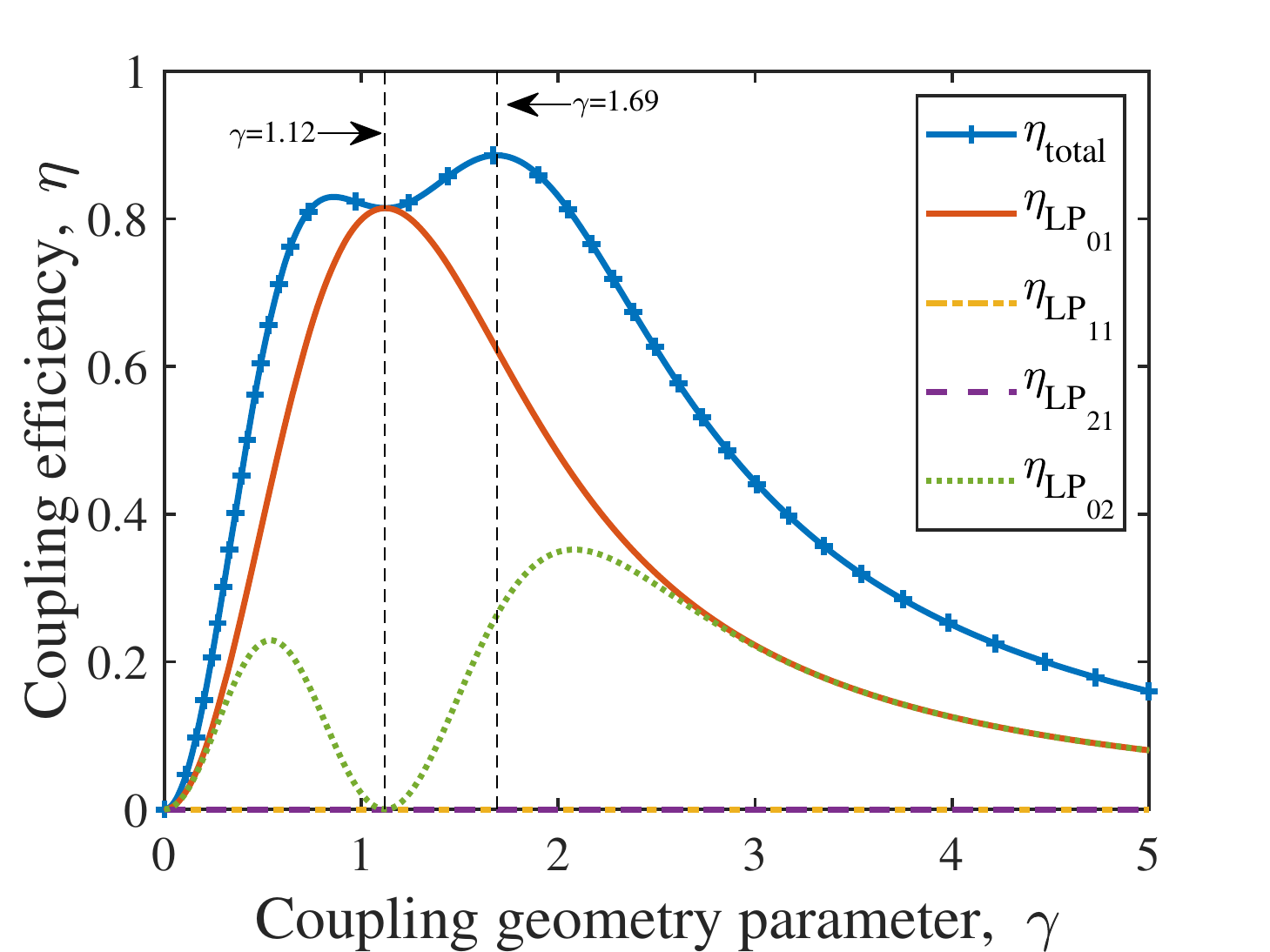}
\caption{Coupling efficiency of each mode from free-space to a six-mode FMF as a function of coupling geometry parameter $\gamma$, in comparison to the total coupling efficiency when the number of speckles $A_R/A_C$ is zero.}
\end{figure}

Figure 3 shows the coupling efficiency of each mode from free-space to a six-mode FMF as a function of coupling geometry parameter $\gamma$, compared to the total coupling efficiency when the number of speckles $A_R/A_C$ is zero (in the absence of turbulence). Similarly, we use $\gamma_6$ to represent the coupling geometry parameter of the six-mode FMF. The coupling efficiency of $\mathrm{LP_{11}}$, $\mathrm{LP_{21}}$ modes are always equal to zero with the change of $\gamma_6$. Since a fiber waveguide energized by a typically incident plane wave, only the $\mathrm{LP}_{0p}$ modes are excited. Through a similar simulation, the parameter $\gamma_6=1.69$ is obtained. 

On the one hand, $\mathrm{LP}_{0p}$ modes for $p\!>\!1$ are excited less efficiently than the $\mathrm{LP_{01}}$ mode, due to the field vectors reverse \cite{snitzer1961cylindrical}. The field distribution has $p\!-\!1$ zero points in the radial direction due to the radial parameter $p$, while the azimuthal direction has no zero points. This distribution causes the light intensity distribution of $\mathrm{LP}_{0p}$ mode, which consists of a centrally located bright spot and $p\!-\!1$ bright rings. Therefore, as the coupling geometry parameter $\gamma$ increases from zero, we can assume $\omega$ is constant, and the lens radius $d_R/2$ increases from zero. Due to the spacing between the central bright spot and the bright ring, the coupling efficiency of the $\mathrm{LP}_{02}$ mode will reasonably appear that fluctuating trend shown in Fig. 3.

\begin{figure}[tbp]
\centering
\subfigure[]{
\includegraphics[width=1.72in]{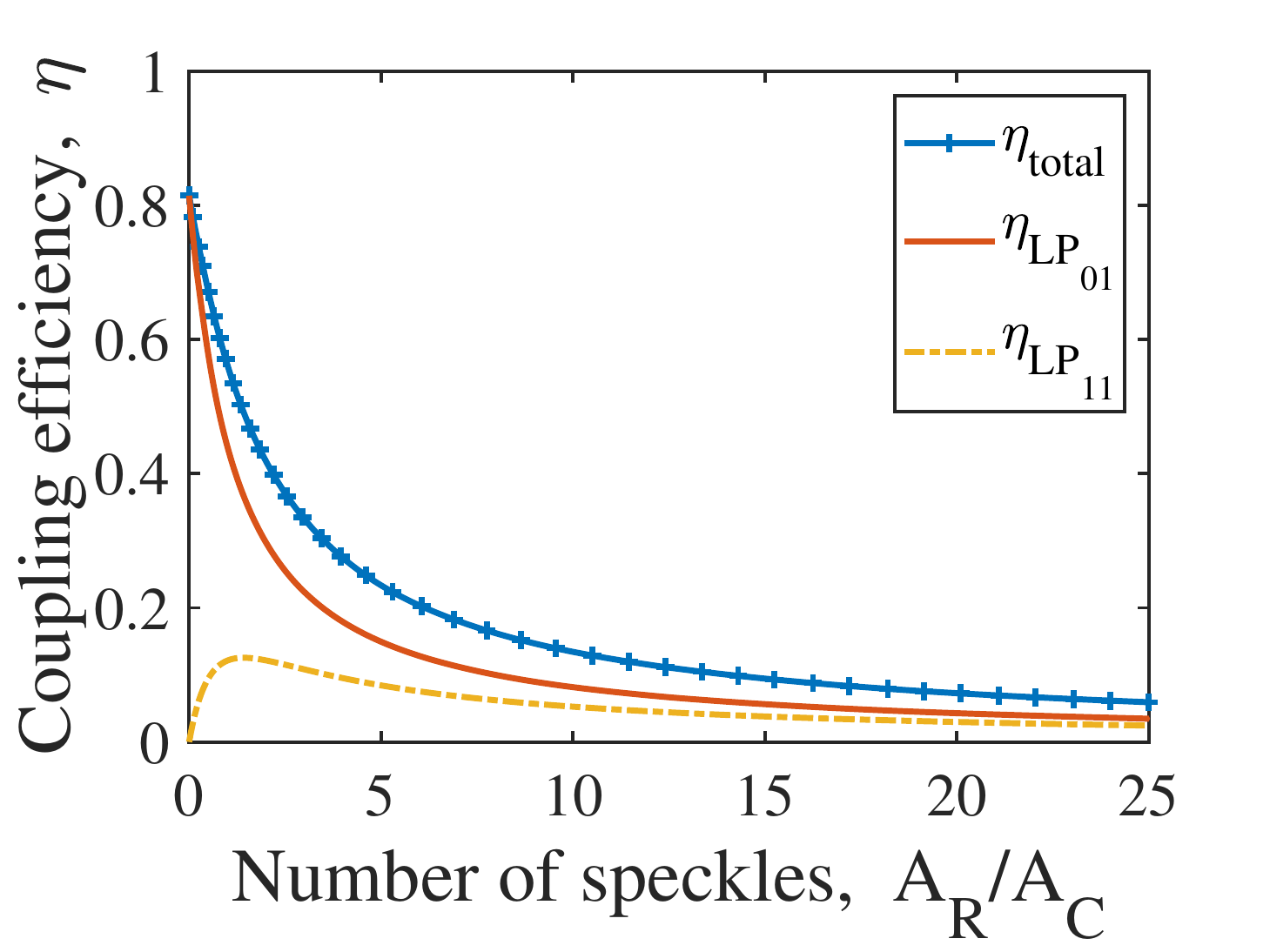}
}
\hspace{-0.24in}
\subfigure[]{
\includegraphics[width=1.72in]{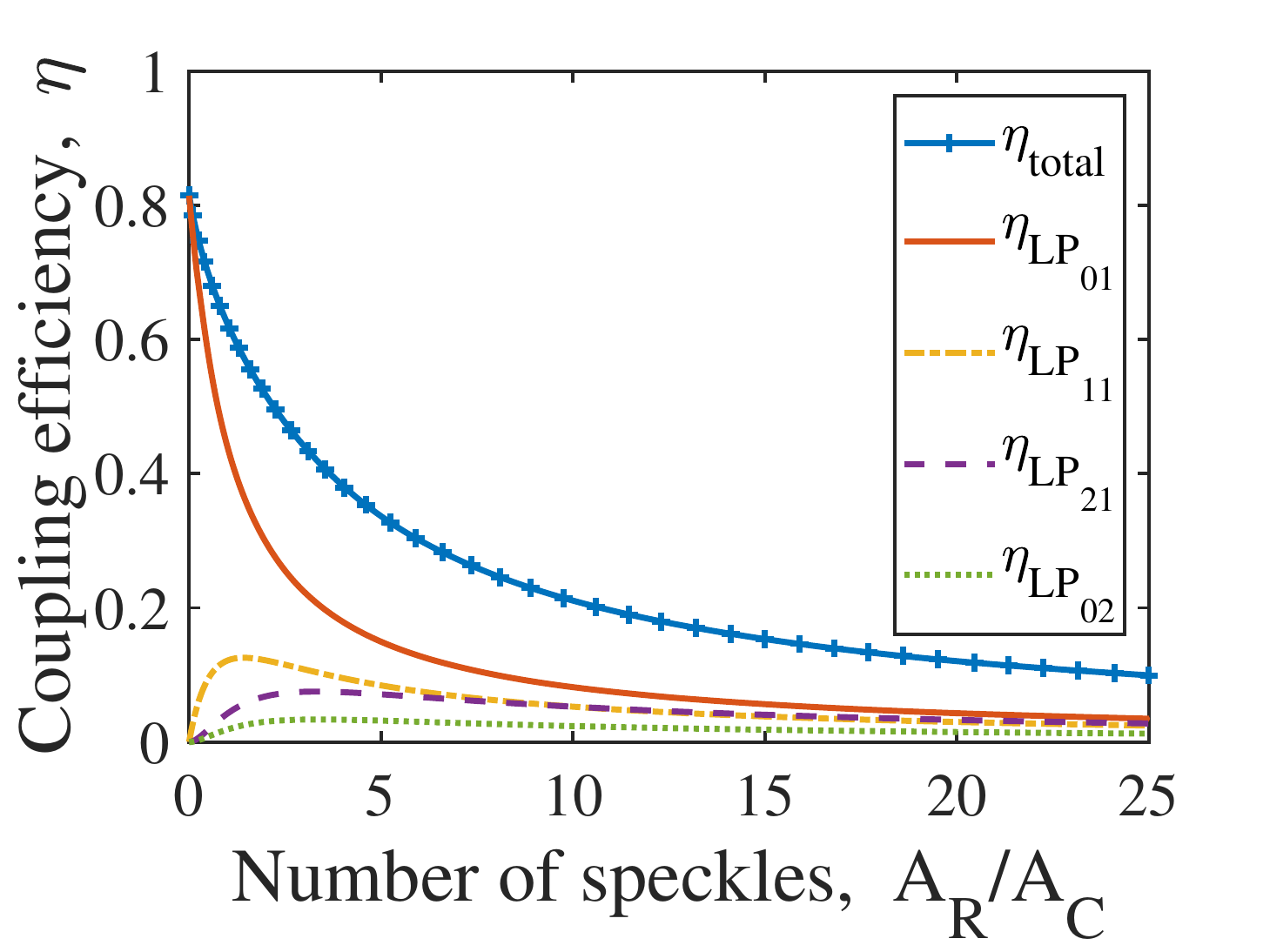}
}
\caption{Coupling efficiency from free-space to FMF as a function of the number of speckles $A_R/A_C$, in comparison to the total coupling efficiency: (a) three-mode free-space to FMF coupling and (b) six-mode free-space to FMF coupling. The coupling geometry parameter $\gamma\!=\!1.12$ applies to both cases.}
\end{figure}

On the other hand, the fundamental mode the $\mathrm{LP_{01}}$ can be well approximated by an untruncated Gaussian function \cite{winzer1998fiber}. Meanwhile, the incident field mutual coherence function given by (\ref{muit}) is also approximated by a Gaussian function. Therefore, we can find that the incident beam wave can be seen as the "transmission mode" in the free-space of the fiber mode whose azimuthal parameter is zero in an ideal case. In other words, the main component of the incident optical wave coupling into the FMF would be the $\mathrm{LP_{01}}$ mode, especially in the absence of turbulence. 

Since the fundamental mode $\mathrm{LP}_{01}$ is the most dominant mode in optical communication, and to facilitate comparison with the three-mode FMF. We set the boundary condition that the initial coupling efficiency of high-order modes approaching minimum value instead of taking the maximum value of the total coupling efficiency, $\gamma_6\!=\!1.12$ is obtained. According to the simulation result, the coupling efficiency of the $\mathrm{LP}_{02}$ mode approaches zero when $\gamma_6\!=\!1.12$. This result is caused by the associated Laguerre polynomials when calculating the integral in (\ref{etajT2}). However, to avoid the inter-mode crosstalk from the $\mathrm{LP}_{02}$ mode, this approximation is acceptable and does not affect our further analysis. Therefore, we use $\gamma\!=\!1.12$ for the remaining calculations of both three-mode and six-mode FMFs.

\subsection{Numerical Results of FMF Coupling Efficiency with Turbulence}
Figure 4(a) shows the three-mode FMF coupling efficiency as a function of the number of speckles $A_R/A_C$ when $\gamma\!=\!1.12$. The coupling efficiency of the $\mathrm{LP_{01}}$ mode decreases rapidly as the number of speckles increases. Especially the number of speckles in the interval of 0 to 10, the coupling efficiency of the $\mathrm{LP_{01}}$ mode has dropped 0.733. Fig. 4(b) shows the six-mode FMF coupling efficiency as a function of the number of speckles $A_R/A_C$ when $\gamma\!=\!1.12$. In this case, the difference between the total coupling efficiency and the coupling efficiency of the fundamental mode is larger than that of the three-mode FMF. With the steady growth of $A_R/A_C$ for both cases above, the coupling efficiency of the $\mathrm{LP_{11}}$ mode undergoes a brief rise and then falls. Since $\mathrm{LP_{11a}}$ mode and $\mathrm{LP_{11b}}$ mode have phase antisymmetry in the absence of turbulence\cite{gloge1971weakly}. When they are superimposed, for a fiber waveguide energized by a normally incident plane wave, the coupling efficiency of the $\mathrm{LP_{11}}$ mode is zero. However, atmospheric turbulence that causes wavefront distortion destroys this phase antisymmetry, which in turn leads to a transient rise in the coupling efficiency of the $\mathrm{LP_{11}}$ mode. When $A_R/A_C=5$, the coupling efficiency of three-mode FMF is 0.234, while that of the $\mathrm{LP_{01}}$ mode is 0.150. It is found that FMF's coupling efficiency is significantly better than that of SMF under the same conditions.

\begin{figure}[tbp]
\centering
\subfigure[]{
\includegraphics[width=1.72in]{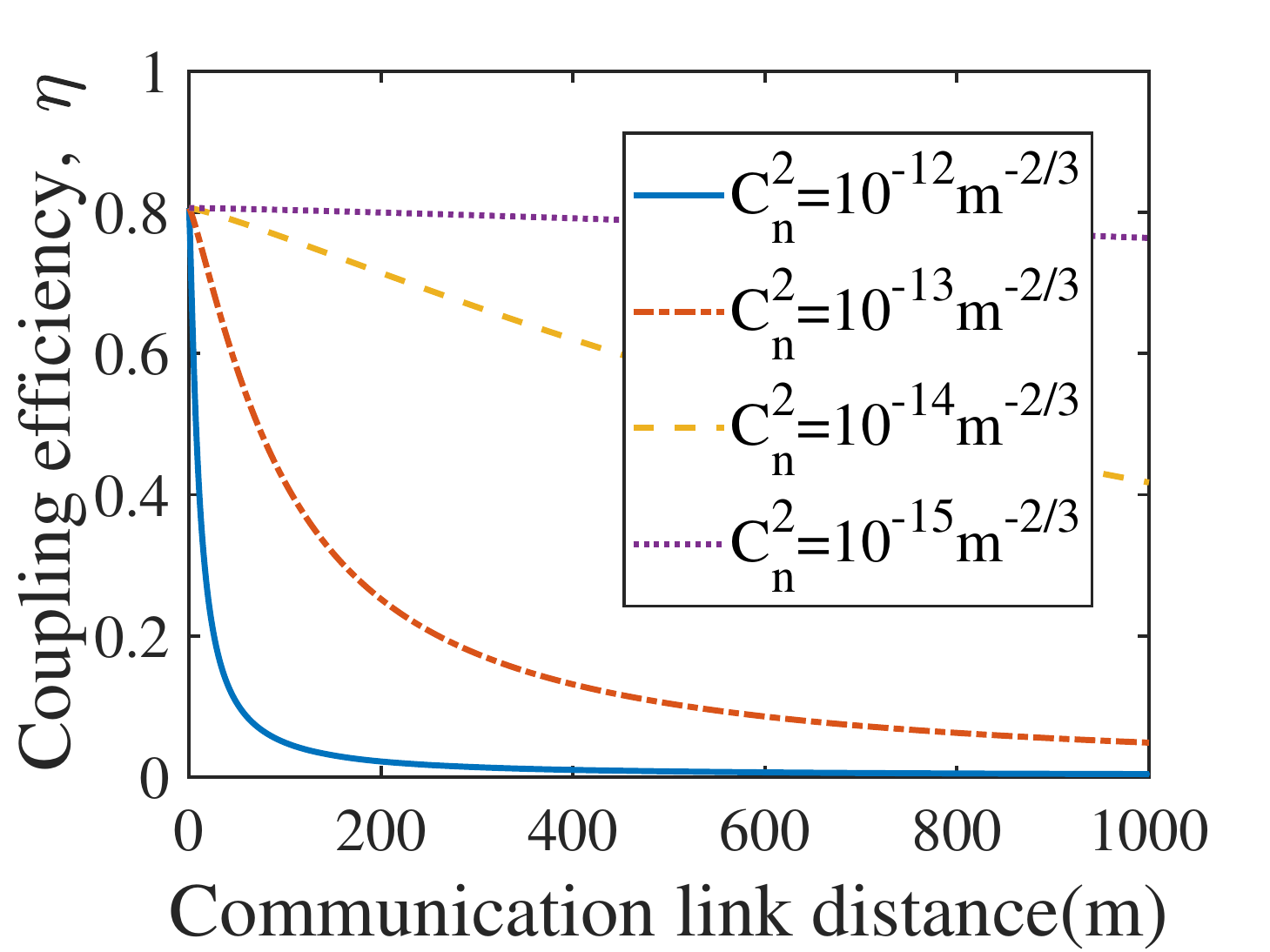}
}
\hspace{-0.24in}
\subfigure[]{
\includegraphics[width=1.72in]{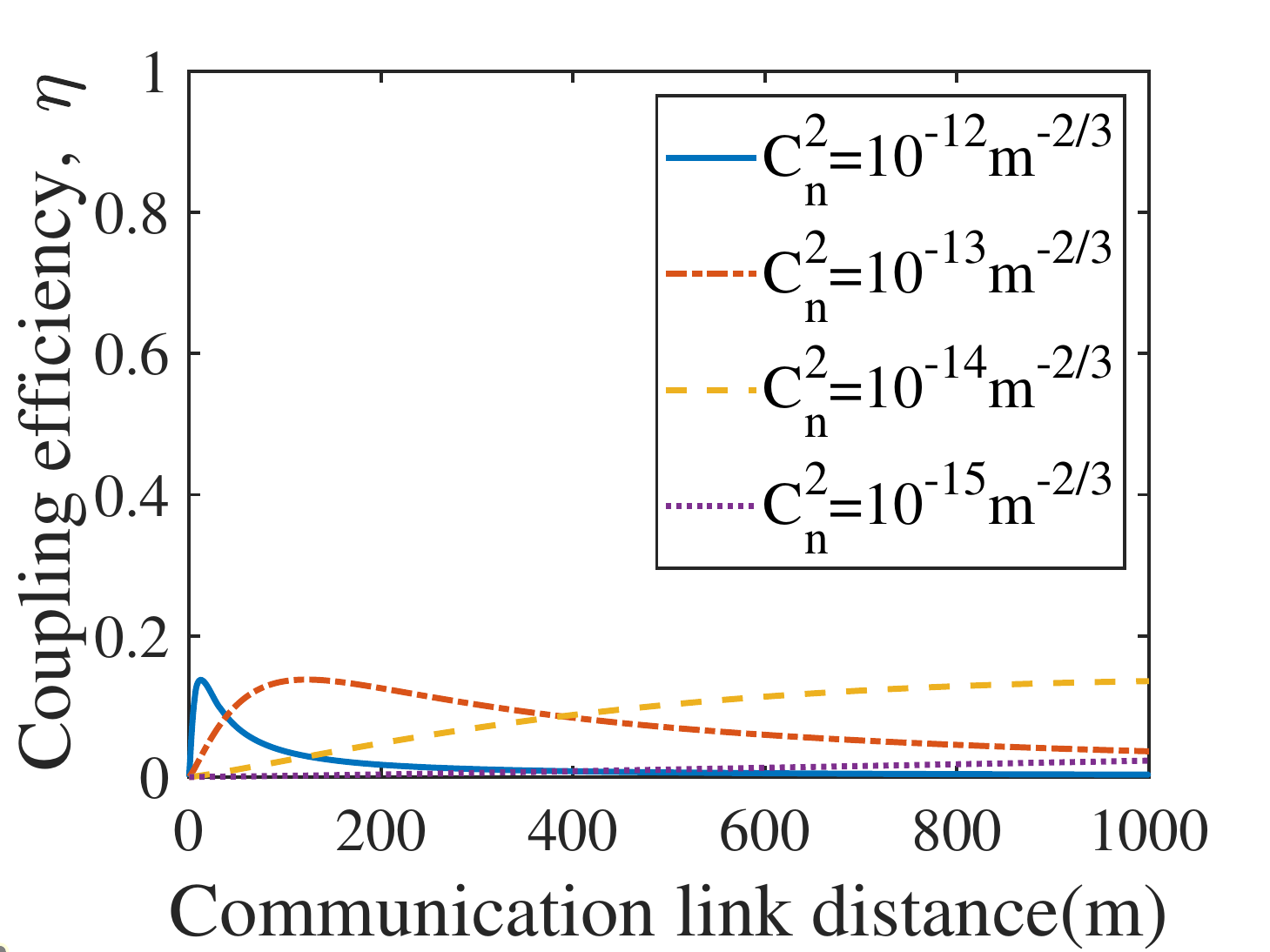}
}
\caption{Coupling efficiency for (a) $\mathrm{LP_{01}}$ mode and (b) $\mathrm{LP_{11}}$ mode as a function of communication link distance with different values of $C_n^2$. The coupling geometry parameter $\gamma\!=\!1.12$ , $d_R=8\,\mathrm{cm}$ and the wavelength is $1550\,\mathrm{nm}$}.
\end{figure}

To understand further the effect of atmospheric turbulence on the coupling efficiency of each mode, Fig. 5(a) and Fig. 5(b) show the coupling efficiency for $\mathrm{LP_{01}}$ mode and $\mathrm{LP_{11}}$ mode as a function of communication link distance with different values of the refractive-index structure constant $C_n^2$. The receiver lens diameter is taken to be $8\,\mathrm{cm}$ and the wavelength is $1550 \,\mathrm{nm}$. We observe values of $C_n^2$ to range from about $10^{-12}$ to $10^{-15} \,\mathrm{m}^{-2/3}$ \cite{tunick2005characterization}. Here we use a value of $C_n^2$, $10^{-12} \,\mathrm{m}^{-2/3}$ to indicate a highly turbulent atmosphere. Alternately, a lower value of $10^{-15} \,\mathrm{m}^{-2/3}$ indicate more adiabatic conditions. The upper and lower curves in Fig. 5(a) can be considered as upper and lower bounds on the coupling efficiency of fundamental mode $\mathrm{LP_{01}}$. A communication link distance of zero yields the maximum coupling efficiency for all different turbulence conditions. Under strong turbulence conditions (where $C_n^2=10^{-12} \,\mathrm{m}^{-2/3}$), the coupling efficiency drops sharply from 0.806 to 0.104 when the distance of the communication link increases to 50$\,\mathrm{m}$.

At the same time, the coupling efficiency of the $\mathrm{LP_{11}}$ mode is different. The coupling efficiency is zero when the communication link distance is zero. Then it experiences a process of rising and falling in all different turbulent conditions. We take the medium value of $C_n^2\!=\!10^{-13} \,\mathrm{m}^{-2/3}$ as an example to analyze the $\mathrm{LP_{11}}$ mode. For a fiber waveguide energized by a normally incident plane wave, $\mathrm{LP_{11}}$ mode is not excited because there is a phase mismatch between the incident beam and one half of the mode pattern. In effect, the mismatch cancels out the excitation received by the other half of the mode pattern. However, when the incident beam is disturbed by atmospheric turbulence, the mode is excited because the cancellation becomes imperfect. When the communication link distance begins to increase from $0$ to about 120$\,\mathrm{m}$, the incident plane wave changes from the ideal state to the disturbed state, and the coupling efficiency increases to the maximum value 0.138. As the distance of the communication link continues to increase, the coupling efficiency begins to decline, and the downward trend is more gradual than the previous upward trend. This trend becomes more sharply with the intensity of atmospheric turbulence increases.

\begin{figure}[tbp]
\centering
\subfigure[]{
\includegraphics[width=1.72in]{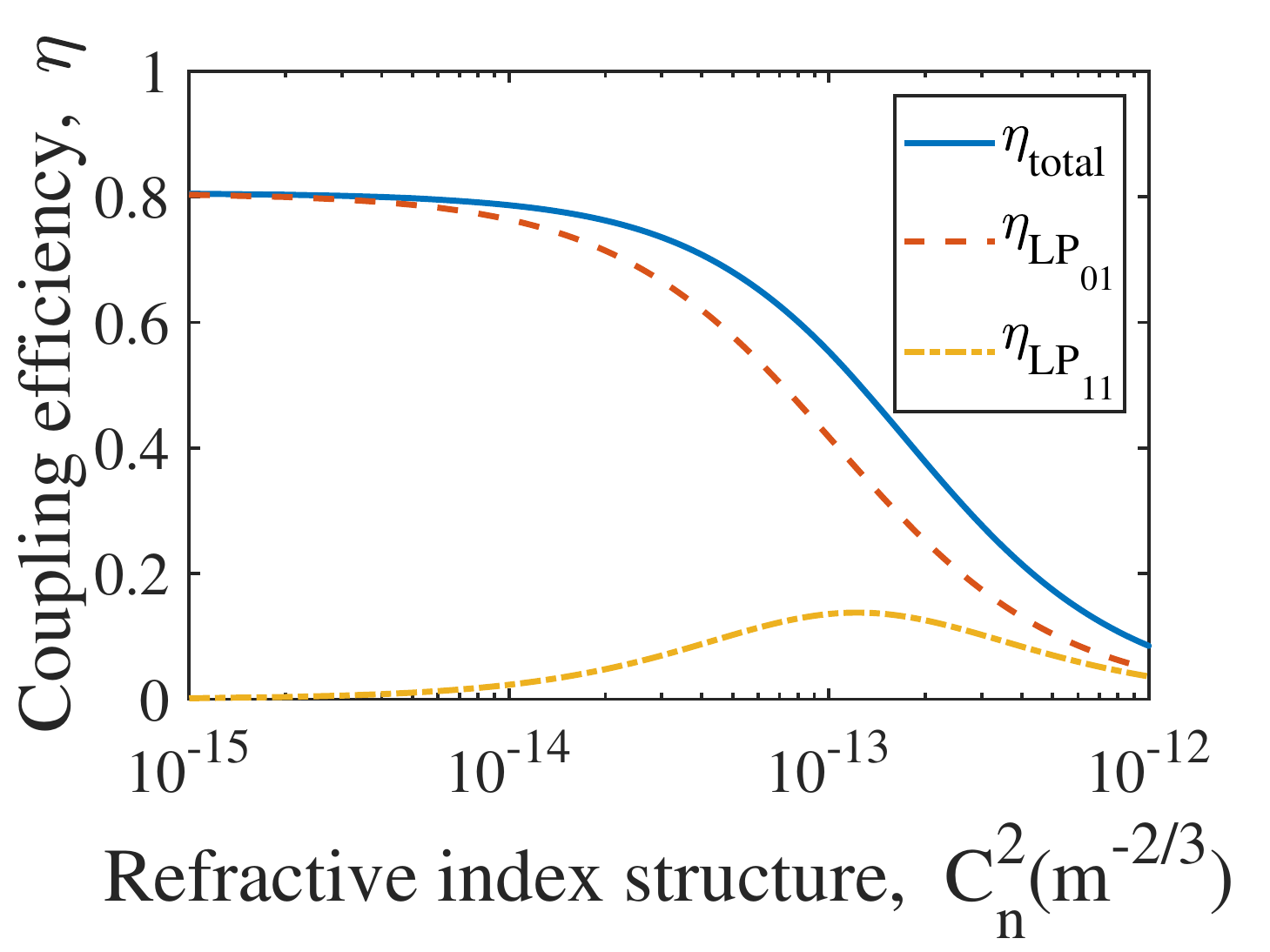}
}
\hspace{-0.24in}
\subfigure[]{
\includegraphics[width=1.72in]{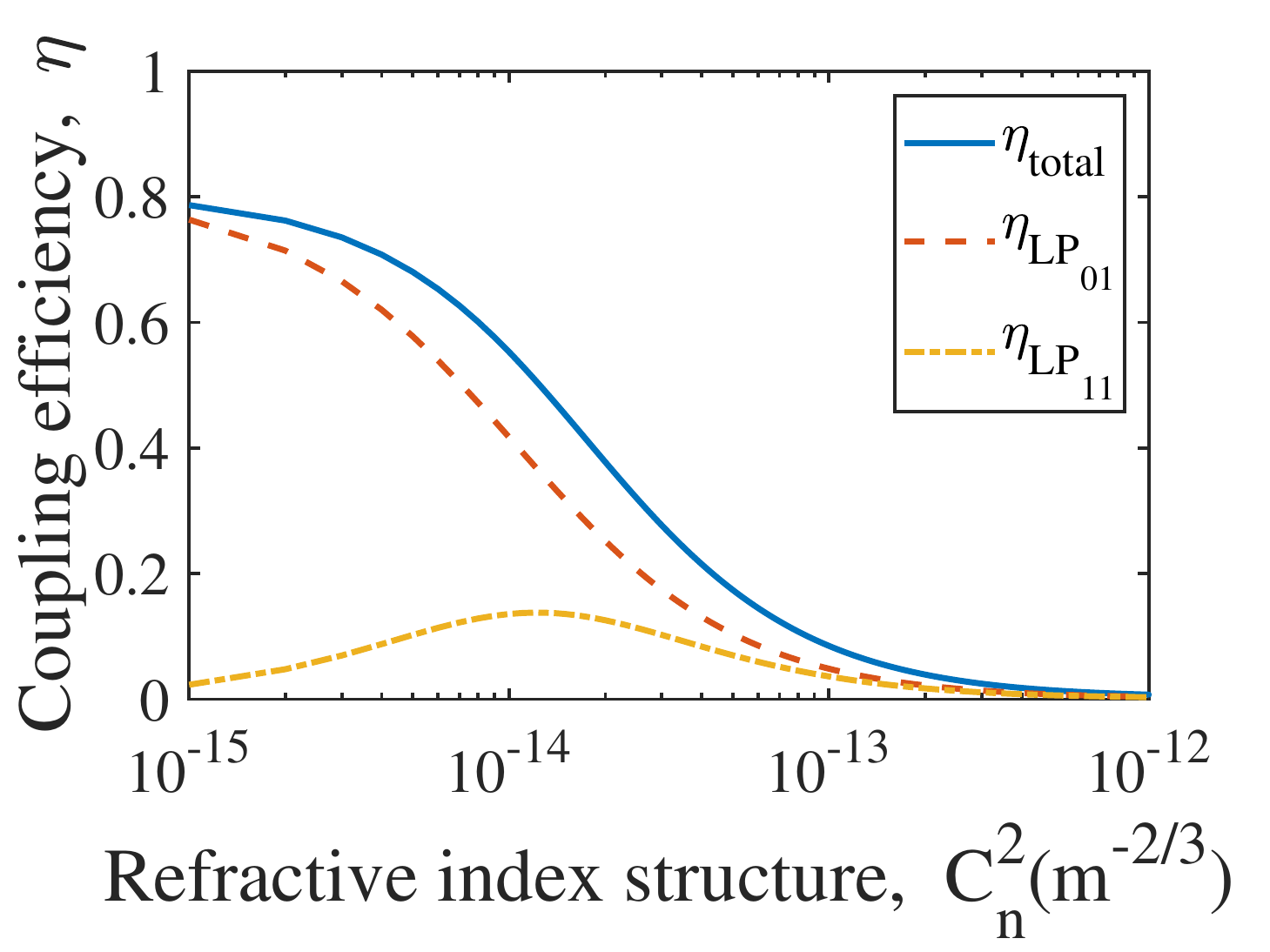}
}
\caption{Coupling efficiency from free-space to a three-mode FMF as a function of $C_n^2$ for a fixed link distance in comparison to the total coupling efficiency. The coupling geometry parameter $\gamma\!=\!1.12$, $d_R=\SI{8}{\centi\metre}$ and the wavelength is $\SI{1550}{\nano\metre}$}. The fixed link distance of (a) is \SI{100}{\metre} and (b) is \SI{1000}{\metre}.
\end{figure}

Figure 6 shows the coupling efficiency from free-space to a three-mode FMF as a function of turbulence strength for a fixed communication link distance. Due to the existence of the $\mathrm{LP_{11}}$ mode, the total coupling efficiency decreases not as rapidly as the coupling efficiency of fundamental mode $\mathrm{LP_{01}}$ with increasing $C_n^2$. For a moderate turbulence strength of $C_n^2\!=\!10^{-13} \,\mathrm{m}^{-2/3}$, the difference value between these two coupling efficiencies is 0.136 as visualized in Fig. 6(a). The result shows that the FMF has better resistance to turbulence than the SMF with the same coupling geometry parameter and communication link distance. When the communication link distance is increased to 1000$\,\mathrm{m}$, as shown in Fig. 6(b), the coupling efficiency experiences a similar intensity change, while this trend is much more sharply. Facing the same turbulence strength as $C_n^2\!=\!10^{-13} \,\mathrm{m}^{-2/3}$, the total coupling efficiency drops to 0.085 when link distance is 1000$\,\mathrm{m}$, while the value for 100$\,\mathrm{m}$ is 0.553.

\subsection{Numerical Results of FMF Coupling Efficiency with Random Jitter}

\begin{figure}[htbp]
\centering
\subfigure[]{
\includegraphics[width=1.72in]{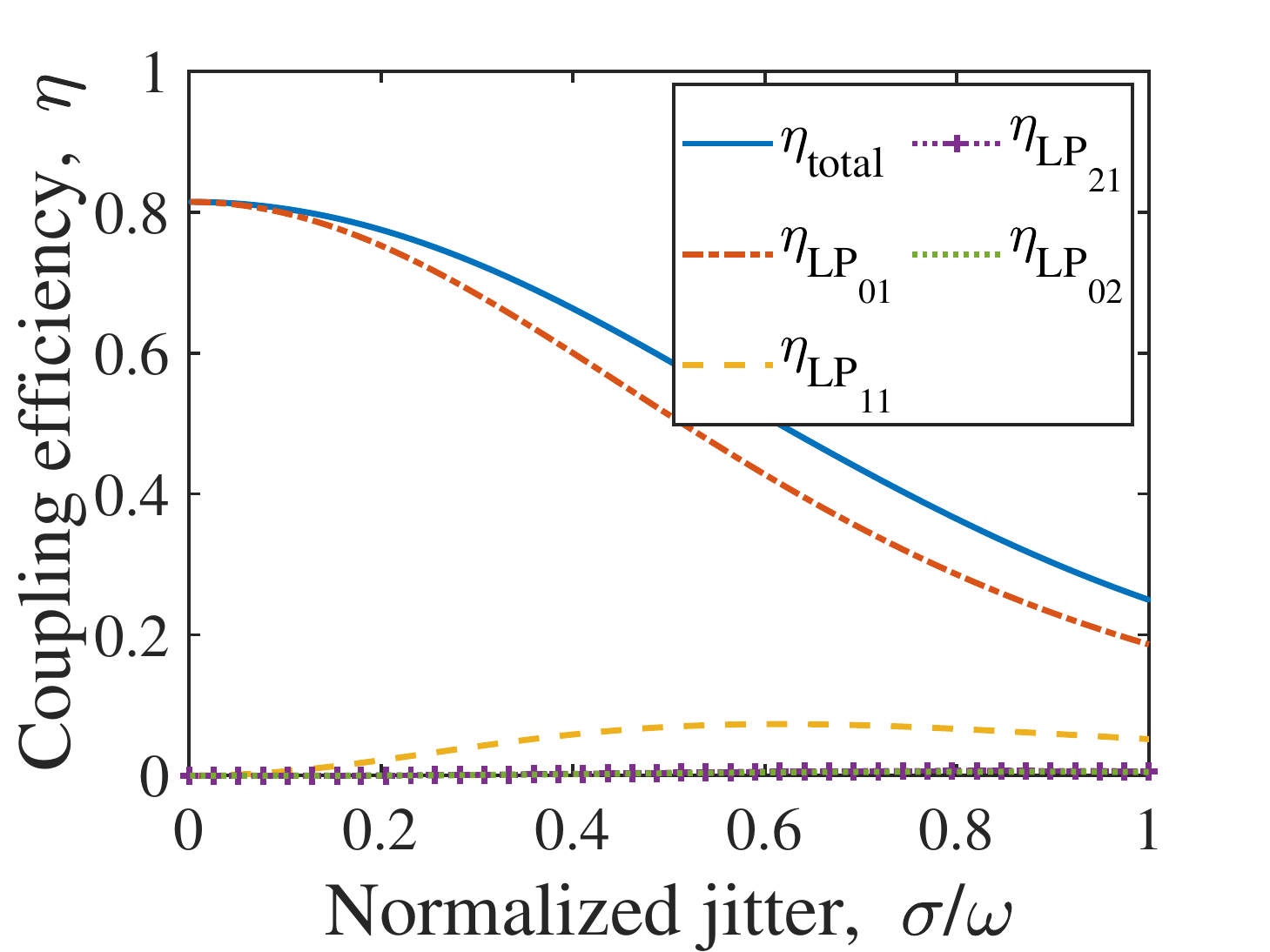}
}
\hspace{-0.24in}
\subfigure[]{
\includegraphics[width=1.72in]{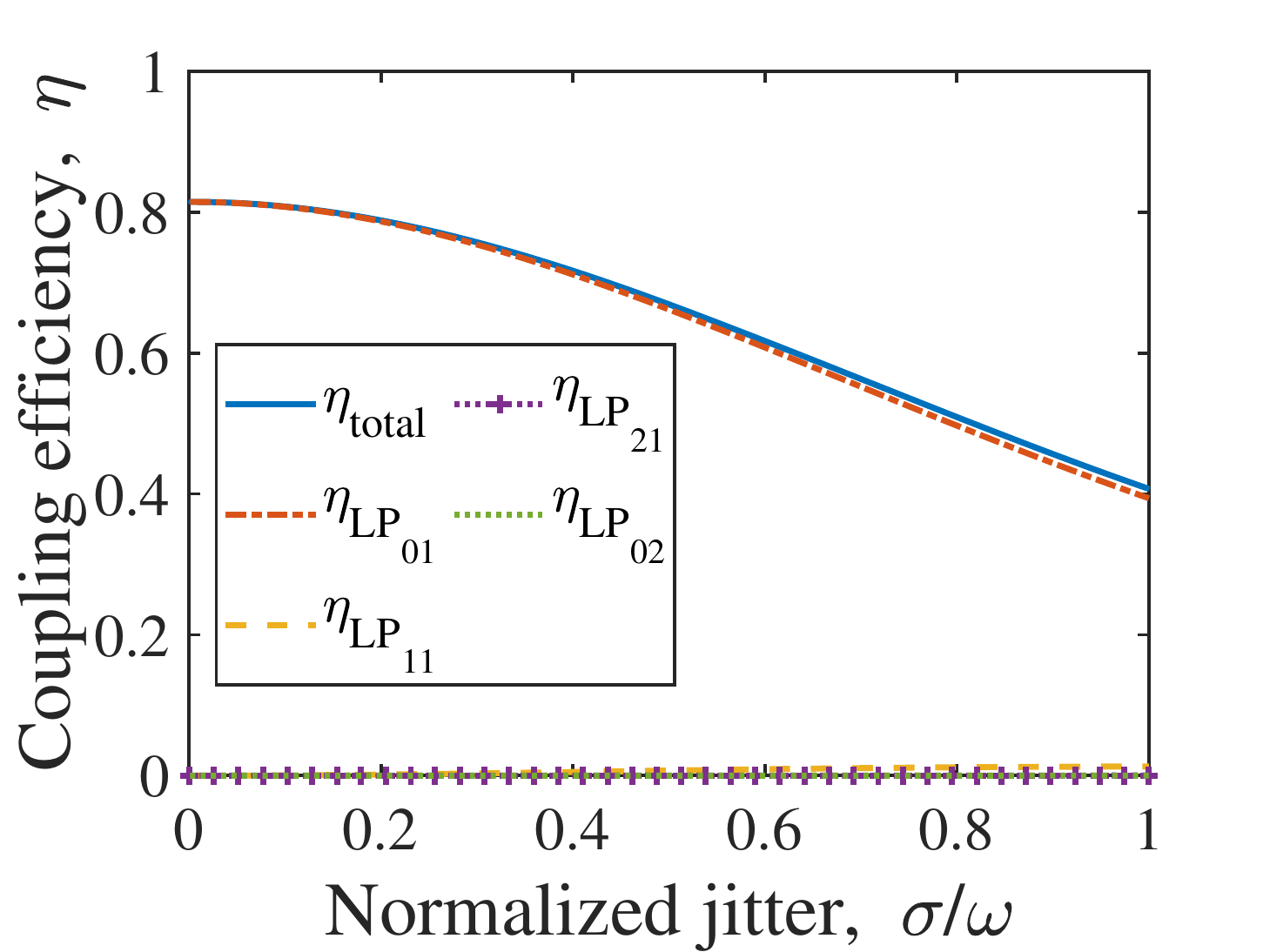}
}
\caption{Coupling efficiency from free-space to a six-mode FMF as a function of random jitter $\sigma$ normalized by the radius of backpropagated fiber mode $\omega$ in comparison to the total coupling efficiency. The receiver lens diameter $d_R=8\,\mathrm{cm}$.} and the coupling geometry parameter $\gamma=1.12$. The relative radius of incident beam $\omega_z/\omega$ for (a) is 1, for (b) is 2.
\end{figure}

Considering the jitter's existence, similarly, we set the receiver lens diameter $d_R=8\,\mathrm{cm}$. Besides, we find that the value of $k/R_z$ has little effect on the coupling efficiency, to simplify our analysis, we can reasonably assume $k/R_z=0$ to obtain the optimum value of $\gamma$. Therefore, eq. (\ref{etajJ2}) can be simplified to the same expression as (\ref{t0}) in the absence of offset bias, and we can obtain the same optimum value of $\gamma=1.12$ in the presence of random jitter. Fig. 7 shows the six-mode FMF coupling efficiency as a function of the normalized jitter $\sigma/\omega$ with different relative radius of incident beam $\omega_z/\omega$. Fig. 7 illustrates that the $\mathrm{LP_{21}}$ mode has low coupling efficiency, while the $\mathrm{LP_{01}}$ mode takes the most of the total coupling efficiency. 

When $\omega_z/\omega$ is changed from 1 to 2\footnote{The parameter $\omega_z$ is defined as the radius at the field amplitudes that fall to $1/e$ at the distance $z$ along the beam. For convenient to discuss, we assume that when $\gamma=1.12$, the incident beam that satisfies $\omega_z/\omega=1$ can completely cover the aperture without random jitter.}, the coupling efficiency of the $\mathrm{LP_{01}}$ mode increases when the normalized jitter is the same. Compared with the $\mathrm{LP_{01}}$ mode, the coupling efficiency of the $\mathrm{LP_{02}}$ mode has a similar conclusion, but its coupling efficiency is too small, the variation tendency is not apparent. Meanwhile, the coupling efficiency of the $\mathrm{LP_{11}}$ mode undergoes a gentle rise and then falls with the normalized jitter in both cases. Similar to the reason for this trend caused by atmospheric turbulence, when the incident beam is disturbed by random jitter, $\mathrm{LP_{11}}$ mode is excited because of the cancellation between the $\mathrm{LP_{11a}}$ mode and the $\mathrm{LP_{11b}}$ mode becomes imperfect until the coupling efficiency peaked \cite{Imai:75}. As random jitter becomes severe, the coupling efficiency begins to decline. However, the coupling efficiency for the $\mathrm{LP_{11}}$ mode tends to be zero as $\omega_z/\omega$ increases, although there are still some impacts on the total coupling efficiency. It can be explained as the increase in the radius of the incident beam would reduce the influence of the jitter on the coupling efficiency. Furthermore, the phase mismatch of the higher-order modes is less affected.

We also find that the coupling efficiency of FMF is slightly better than that of SMF (i.e., $\mathrm{LP_{01}}$ mode) in the presence of random jitter under the same conditions, where the effect of core radius on coupling efficiency is ignored.

\subsection{Communication Performance with Random Jitter}

{
\begin{figure}[t]
        \centering
        \subfigure[]{
\includegraphics[width=1.72in]{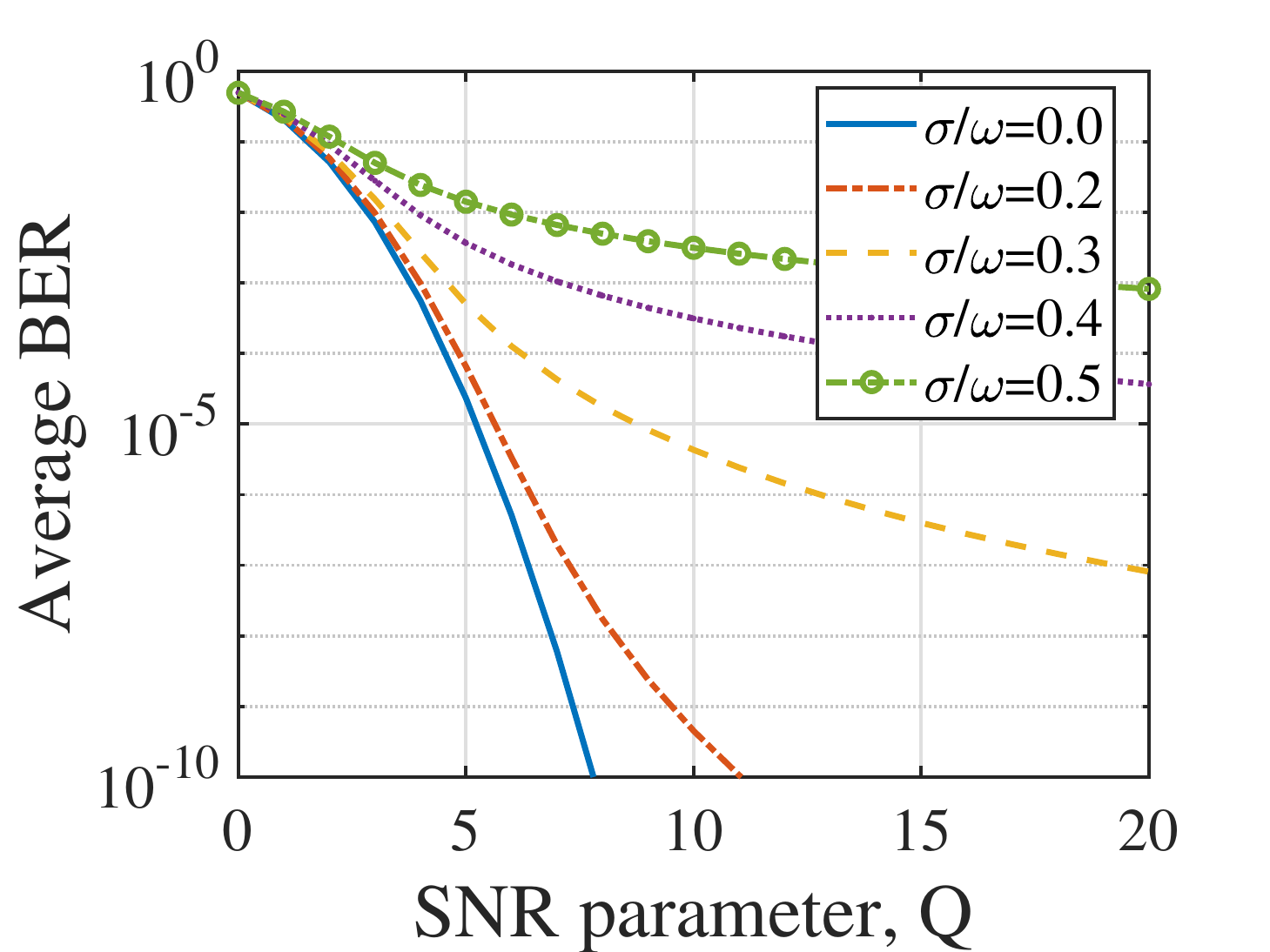}
}
\hspace{-0.24in}
\subfigure[]{
\includegraphics[width=1.72in]{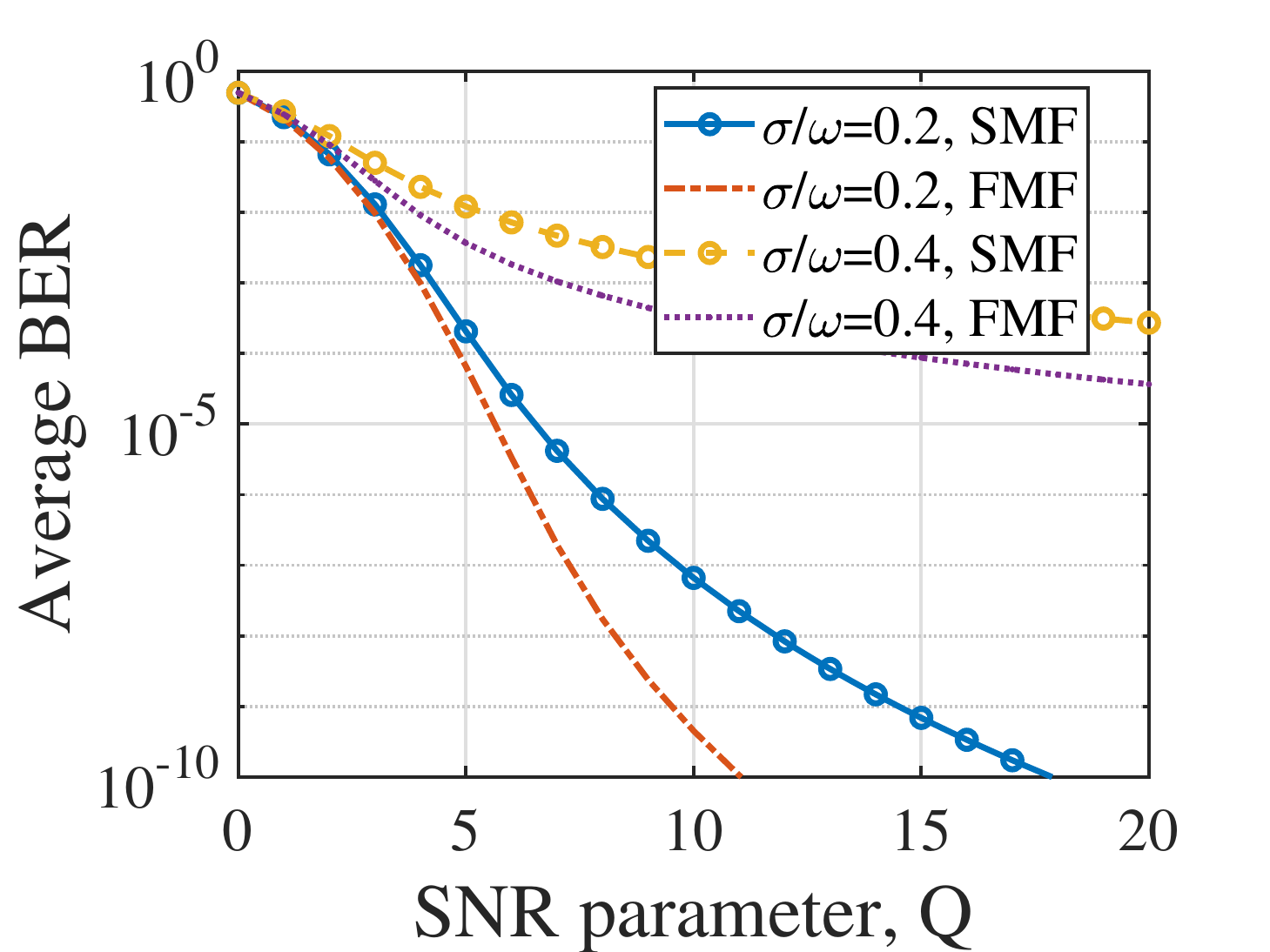}
}
\caption{Average BER as a function of normalized jitter and the SNR parameter. (a) six-mode FMF in different normalized jitter condition and (b) six-mode FMF compared with SMF in same normalized jitter condition. For both cases, the receiver lens diameter $d_R=8\,\mathrm{cm}$, the coupling geometry parameter $\gamma=1.12$ and the relative radius of incident beam $\omega_z/\omega=1$.}
\end{figure}

Figure 8 shows the average BER as a function of random jitter. We can find that for six-mode FMF, the average BER degrades significantly when the random jitter $\sigma/\omega>0.3$. Compared with SMF, the six-mode FMF has better BER performance in the presence of random jitter. Besides, a smaller value of random jitter can make the BER performance of FMF have a greater improvement than that of SMF in high SNR regimes.}

\section{CONCLUSION}
We proposed a theoretical coupling model from the FSO link to FMF, which is based on a scale-adapted set of LG modes. This set of LG modes can well approximate the LP modes of a step-index fiber. It is found that the coupling efficiency from the FSO link to FMF of different modes behaves differently in the presence of atmospheric turbulence or random jitter. Specifically, taking a six-mode FMF as an example, the coupling efficiency of $\mathrm{LP}_{0p}$ modes drops as turbulence or random jitter becomes more severe, and the $\mathrm{LP}_{02}$ mode is excited less efficiently than the $\mathrm{LP_{01}}$ mode, due to the field vectors reverse. At the same time, the coupling efficiency of the $\mathrm{LP}_{11}$ mode reaches the peak then drops rapidly, which is caused by the phase mismatch between the incident beam and one half of the mode pattern. However, the total coupling efficiency will decline with turbulence intensity or random jitter. We also deduced the optimum value for the coupling geometry parameter $\gamma$ to maximize the ratio of the coupling efficiency of the fundamental mode to the total coupling efficiency, for both three-mode FMF and six-mode FMF. Correspondingly, we compared the influence of different parameters on each mode's coupling efficiency with random jitter. As a comparison, the coupling efficiency of the $\mathrm{LP_{01}}$ mode can be approximated as the coupling efficiency of SMF. We found that FMF's coupling efficiency is better than that of SMF to varying degrees under the same conditions, whether it is turbulence or random jitter. Finally, the communication performance of six-mode FMF with random jitter is investigated. The average BER degrades significantly when the random jitter $\sigma/\omega>0.3$. Compared with SMF, the FMF performs better, especially in high SNR regimes. Our future work will consider the effects of more complicated situations, including both turbulence and random jitter on the coupling efficiency of FMF.


\end{document}